\documentclass[aps, prl, reprint, superscriptaddress]{revtex4-2}
\usepackage{amsmath}
\usepackage{amssymb}
\usepackage{amsfonts}
\usepackage{mathrsfs}
\usepackage{booktabs, array}
\usepackage[T1]{fontenc}
\usepackage[caption=false]{subfig}
\usepackage[dvipsnames=true]{xcolor}
\usepackage{tikz}
\usepackage{graphicx} 
\usepackage{multirow}
\usepackage{makecell}

\definecolor{linkcolor}{rgb}{0.0,0.3,0.5}
\usepackage[
    hypertexnames=false,
    unicode,
    colorlinks=true,
    linkcolor=linkcolor,
    citecolor=linkcolor,
    filecolor=linkcolor,
    urlcolor=linkcolor,
    pdfusetitle
]{hyperref}
\usepackage{orcidlink}

\newcommand{\UIUC}{Illinois  Center  for  Advanced  Studies  of  the  Universe \&
Department of Physics, University of Illinois at Urbana-Champaign, Urbana, Illinois 61801, USA}

\graphicspath{{./Plots/}}
\begin{document}
\title{From the Test-Mass Limit to Binary Black-Hole Waveforms \\
in Higher-Derivative Gravity}
\author{Chaoyi Yang }
\affiliation{Department of Astronomy, Tsinghua University, Beijing 100084, China}

\author{Neev Khera \orcidlink{0000-0003-3515-2859}}
\affiliation{Department of Astronomy, Tsinghua University, Beijing 100084, China}

\author{Dongjun Li \orcidlink{0000-0002-1962-680X}}
\thanks{Contact author: \href{dongjun@illinois.edu}{dongjun@illinois.edu}}
\affiliation{\UIUC}

\author{Huan Yang \orcidlink{0000-0002-9965-3030}}
\thanks{Contact author: \href{hyangdoa@tsinghua.edu.cn}{hyangdoa@tsinghua.edu.cn}} 
\affiliation{Department of Astronomy, Tsinghua University, Beijing 100084, China}

\begin{abstract}
Many higher-derivative theories predict stronger deviations from General Relativity for lower-mass black holes, while their nonlinear field equations often prevent reliable simulations of the full binary evolution. Here we develop a route from controlled black hole perturbation theory based on the modified Teukolsky formalism to comparable-mass waveforms, using parity-even cubic gravity as a representative example. We find that the tidal response of the secondary black hole enters at the same perturbative order as the direct higher-curvature correction and is therefore essential for a consistent leading-order waveform. The resulting strong-field fluxes and conservative dynamics produce an accumulated inspiral dephasing that grows toward merger. Embedding this test-mass information into an effective-one-body model, we construct inspiral-merger-ringdown waveforms for comparable-mass binaries and find coupling-dependent dephasing and waveform-peak shifts. Our results demonstrate how strong-field test-mass calculations can anchor waveform models for higher-derivative gravity when theory-specific numerical-relativity simulations are unavailable.
\end{abstract}
\maketitle

\noindent{{\emph{Introduction}}---Extreme mass-ratio inspirals (EMRIs), in which a stellar-mass compact object slowly inspirals into a massive black hole, are among the principal targets of future space-based gravitational-wave observatories \cite{LISA:2022kgy,TianQin:2020hid,Ruan:2018tsw}. Their small mass ratios separate the orbital and radiation-reaction timescales, allowing the secondary to remain in the strong-field region for many cycles. Consequently, weak environmental perturbations can accumulate into sizable waveform dephasings \cite{Kocsis:2011dr,Barausse:2014tra,Copparoni:2025jhq,speri2023measuring,Bonga:2019ycj,Khalvati:2024tzz,Li:2025ffh,Dyson:2025dlj,Zhang:2018kib,Zhang:2019eid,Sun:2025lbr,LaHaye:2025ley}.

The same phase coherence makes EMRIs natural probes of physics beyond General Relativity (GR). Deviations from GR may modify the black hole geometry, wave generation and propagation, or the charges and finite-size response of compact objects. The preferred black hole mass for the strongest observational signals is theory dependent. Local higher-derivative corrections generally scale as positive powers of $\ell/M$, favoring stellar-mass binaries~\cite{Endlich:2017tqa,Cardoso:2018ptl,Cano:2019ore}, and have motivated direct gravitational-wave constraints using stellar-mass binary black holes~\cite{Liu:2024atc, Perkins:2021mhb}. EMRIs may nevertheless remain competitive through radiation sourced by the secondary \cite{Maselli:2020zgv,Barsanti:2022vvl,Yunes:2011aa,Pani:2011xj}, while cosmologically induced hair \cite{Jacobson:1999vr,Babichev:2025ric}, scalarization confined to a supermassive-black-hole mass window \cite{Eichhorn:2023iab,Thaalba:2025ljh}, and infrared modifications enhanced at low frequencies \cite{Will:1997bb,Berti:2004bd,Mirshekari:2011yq} can instead favor massive black holes. A source-driven perturbative framework for massive black holes is therefore relevant to a broad class of beyond-GR theories.

Waveforms are difficult to obtain in many modified-gravity theories, particularly higher-curvature theories whose nonlinear field equations do not yet have a generally applicable well-posed evolution formulation \cite{Delsate:2014hba, Cayuso:2017iqc, Ripley:2019hxt}. Black hole perturbation theory offers an alternative that retains strong-field information without a weak-field expansion. The modified Teukolsky formalism (MTF) extends the Teukolsky description of black hole perturbations to deformed backgrounds and modified field equations \cite{Li:2022pcy, Hussain:2022ins, Cano:2023tmv, Wagle:2023fwl, Li:2023ulk, Li:2025fci, Aly:2026otj}. In this Letter and the companion Article \cite{Yang:2026erj}, we develop the source-driven MTF at first order in both the beyond-GR coupling and the mass ratio. It consistently incorporates the deformed background, modified wave generation and propagation, the orbiting body, and its theory-dependent finite-size response, while avoiding the full nonlinear two-body problem. The same strategy can be extended by coupling the MTF to perturbations of additional dynamical fields \cite{Wagle:2023fwl, Li:2025fci}.

As a concrete application, we study circular inspirals into a nonspinning black hole in parity-even cubic gravity \cite{Stelle:1977ry, Cano:2019ore}. We compute the leading corrections to the gravitational-wave fluxes at null infinity and through the horizon, including both the bulk higher-curvature contribution and the induced quadrupolar response of the secondary black hole. These effects enter at the same order in the coupling and mass ratio, so the tidal response is required for a complete leading-order waveform. Combining the modified fluxes with the conservative dynamics, we obtain the accumulated inspiral dephasing. Although EMRIs are not the optimal systems for constraining this theory, they provide controlled strong-field test-mass information. We use this information, together with finite-mass-ratio post-Minkowskian (PM) dynamics and theory-specific quasinormal modes, to construct effective-one-body (EOB) inspiral--merger--ringdown waveforms for comparable-mass binaries. Representative equal-mass waveforms are shown in Fig.~\ref{fig:imr_waveforms}, illustrating the coupling-dependent dephasing and the extension of the construction through merger and ringdown. This establishes a route from source-driven black hole perturbation theory to waveform models for the stellar-mass systems most sensitive to higher-derivative gravity.

\begin{figure*}[t]
    \centering
    \includegraphics[width=\textwidth]
    {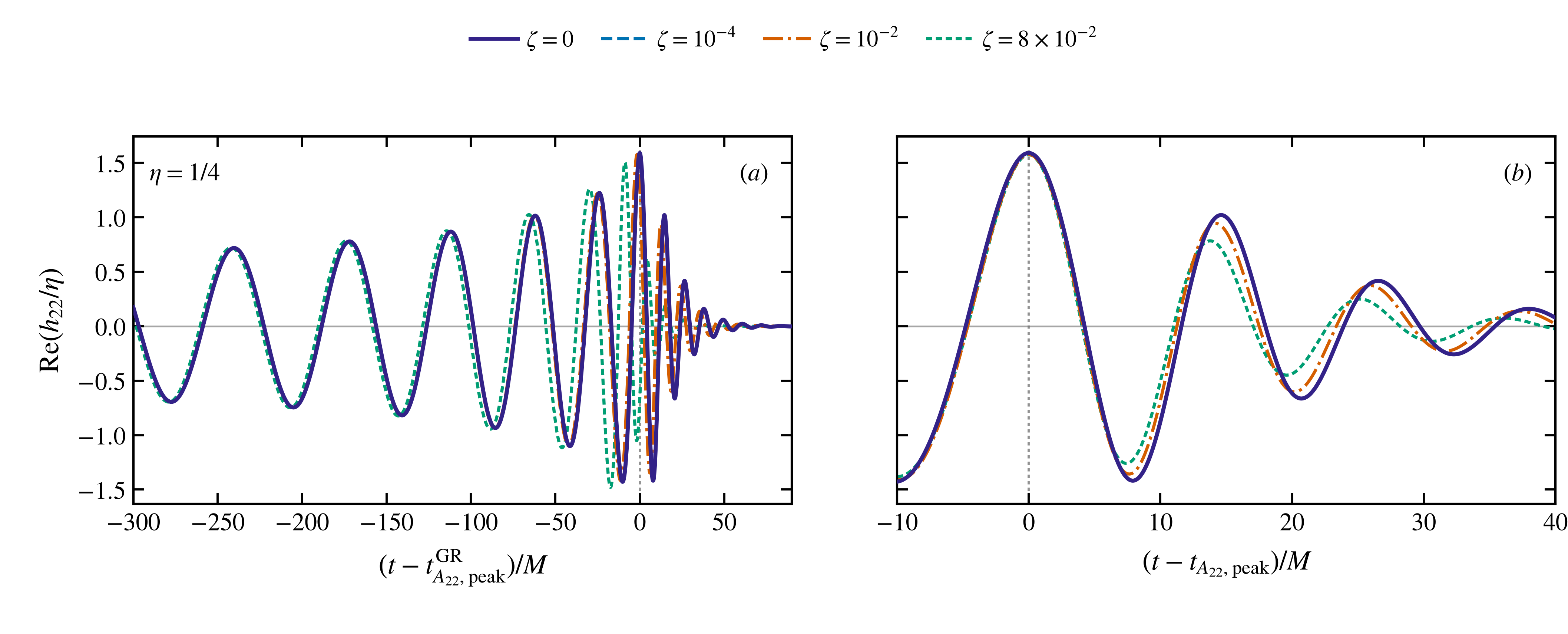}
    \caption{The real part of the normalized $(2,2)$ mode, $\mathrm{Re}(h_{22}/\eta)$, for an equal-mass binary ($\eta=1/4$) at different values of $\zeta$. (a) Complete inspiral-merger-ringdown waveforms shown relative to the GR waveform-amplitude peak, $(t-t_{A_{22},{\rm peak}}^{\rm GR})/M$. All evolutions start at the common reference frequency $M\Omega=20^{-3/2}$ with $t=0$, and the solid blue curve denotes GR ($\zeta=0$). (b) Waveforms near merger and ringdown aligned in time and phase at their individual waveform-amplitude peaks, $(t-t_{A_{22},{\rm peak}})/M$. No amplitude rescaling is applied.
    }
     \label{fig:imr_waveforms}
\end{figure*}

We use units with $G=c=1$.

\vspace{0.2cm}

\noindent{{\emph{Modified Teukolsky formalism}}}---Consider an EMRI within a modified theory of gravity. Compared to GR, the system may feature several important distinctions that can leave measurable imprints on the gravitational waveform. First, the unperturbed background may differ from Kerr because of modifications to the Einstein-Hilbert action. Second, the theory may contain additional dynamical fields, e.g., dynamical Chern-Simons (dCS) theory \cite{Alexander:2009tp} and Einstein-dilaton-Gauss-Bonnet (EdGB) gravity \cite{Pani:2009wy}. Neither feature is universal, but wave generation and propagation in these theories generally differ from those in GR and are essential for determining the long-term evolution of an EMRI. For simplicity, we focus here on pure-metric higher-derivative theories \cite{Stelle:1977ry,Cano:2019ore}; extensions with additional dynamical fields require a coupled perturbative treatment.

The MTF was originally developed to study black-hole quasinormal modes in modified gravity~\cite{Li:2022pcy,Hussain:2022ins,Cano:2023tmv,Wagle:2023fwl,Li:2023ulk,Li:2025fci,Aly:2026otj} and has since been extended to source-driven perturbations relevant to EMRIs in astrophysical environments~\cite{Li:2025ffh,LaHaye:2025ley}. Within the MTF, Newman-Penrose quantities are organized in a two-parameter expansion:
\begin{align}
\Psi &= \Psi^{(0,0)}+\zeta\Psi^{(1,0)}+\eta\Psi^{(0,1)}
+\zeta\eta\Psi^{(1,1)}+\cdots.
\end{align}
Here, $\zeta$ characterizes deviations from GR, and $\eta$ controls the amplitude of the dynamical perturbation. For EMRIs, $\eta$ may be identified with the mass ratio $m_p/M_{\rm BH}\ll1$, where $M_{\rm BH}$ and $m_p$ denote the masses of the primary black hole and its companion, respectively. The superscript $(i,j)$ labels the perturbative order $O(\zeta^i,\eta^j)$. The leading beyond-GR correction to the radiative degrees of freedom is therefore encoded in $\Psi_{0,4}^{(1,1)}$.

For the particle-sourced problem considered here, we organize the mixed-order MTF equations as
\begin{align}
H_{0}^{\mathrm{GR}}\Psi_{0}^{(1,1)}&=\mathcal{S}_{\mathrm{geo}}^{(1,1)}+\mathcal{S}_A^{(1,1)}+\mathcal{S}_B^{(1,1)}\,,\\
H_{4}^{\mathrm{GR}}\Psi_{4}^{(1,1)}&=\mathcal{T}_{\mathrm{geo}}^{(1,1)}+\mathcal{T}_A^{(1,1)}+\mathcal{T}_B^{(1,1)}\,.
\end{align}
Here, $H_{0,4}^{\mathrm{GR}}$ are the standard GR Teukolsky operators.
The terms $\mathcal{S}_{\mathrm{geo}}^{(1,1)}$ and
$\mathcal{T}_{\mathrm{geo}}^{(1,1)}$ arise from the deformation of the Teukolsky operators induced by the modified background, $\mathcal{S}_{A}^{(1,1)}$ and $\mathcal{T}_{A}^{(1,1)}$ arise from the higher-curvature corrections to the Einstein equations, and $\mathcal{S}_{B}^{(1,1)}$ and $\mathcal{T}_{B}^{(1,1)}$ arise from the particle source. The explicit source terms for the cubic-gravity EMRI considered below are given in Ref.~\cite{Yang:2026erj}.

As a working example, we consider parity-preserving cubic gravity, for which the EMRI system is effectively described by
\begin{align}
\begin{split}
     S=&\frac{1}{16\pi}\int d^4x\,\sqrt{|g|}\left(R+\lambda_{\rm ev}l_c^4 
    R_{\mu\nu}{}^{\rho\sigma}
R_{\rho\sigma}{}^{\alpha\beta}
R_{\alpha\beta}{}^{\mu\nu}\right)\\&-m_p\int d\tau\,+\frac{c_E}{2}\int d\tau\,E_{\mu\nu}E^{\mu\nu}
    +\frac{c_B}{2}\int d\tau\,B_{\mu\nu}B^{\mu\nu}.    
\end{split}
\end{align}
Here, the $R_{\mu\nu}{}^{\rho\sigma} R_{\rho\sigma}{}^{\alpha\beta} R_{\alpha\beta}{}^{\mu\nu}$ operator defines the parity-preserving cubic-gravity correction, whose strength is characterized by the dimensionless coupling $\zeta\equiv\lambda_{\rm ev}l_c^4/M_{\rm BH}^4$.
The remaining worldline operators encode the leading quadrupolar tidal response \cite{Wang:2026qst}:
\begin{align}
    & E_{\mu\nu}= C_{\mu\alpha\nu\beta}\,u^\alpha u^\beta ,\\
    & B_{\mu\nu}=\frac{1}{2}\,\epsilon_{\mu\alpha\rho\sigma}\,C^{\rho\sigma}{}_{\nu\beta}\,u^\alpha u^\beta .
\end{align}
Here, $C_{\mu\nu\rho\sigma}$ is the Weyl tensor, $u^\mu$ is the particle four-velocity, and $\epsilon_{\mu\nu\rho\sigma}$ is the Levi-Civita tensor. The tensors $E_{\mu\nu}$ and $B_{\mu\nu}$ are the electric- and magnetic-type tidal fields, respectively. Black holes have a nonvanishing tidal response in this theory, with Wilson coefficients $c_E=28m_p\lambda_{\rm ev}l_c^4/3$ and $c_B=-20m_p\lambda_{\rm ev}l_c^4/3$ fixed by the matching calculations of Refs.~\cite{Cano:2026hlv,Wang:2026qst}.

The cubic interaction is special in that $c_{E,B}\propto m_p\lambda_{\rm ev}l_c^4$. Consequently, the finite-size tidal source and the direct cubic-curvature correction both enter at $O(\zeta\eta)$ and contribute to the energy flux at $O(\zeta\eta^2)$. Our direct cubic-curvature correction to the flux at null infinity agrees in the weak-field regime with the leading 5PN result of Ref.~\cite{Cano:2026hlv}, in which the correction is proportional to the tidal deformability of the primary black hole. A detailed comparison is given in the companion Article~\cite{Yang:2026erj}. Together with the finite-size response of the secondary, this agreement suggests that the leading corrections from the two bodies may share a common tidal origin. More generally, for a pure-metric interaction $\alpha_n R^n$, dimensional matching gives $c_{E,B}^{(n)}\sim\alpha_n m_p^{7-2n}$, so the tidal response is subleading in $\eta$ relative to the direct curvature correction for $n=2$, enters at the same order for $n=3$, and is formally enhanced for $n>3$. Scalar-curvature theories such as dCS and EdGB instead obey a different counting: the metric response is quadratic in the coupling $\alpha$, giving $c_{E,B}^{\rm grav}\sim m_p\alpha^2$, so both the gravitational finite-size and direct gravitational-flux corrections enter at $O(\alpha^2\eta^2)$. Scalar radiation can enter at lower order in the mass ratio: for a secondary with fixed dimensionless spin in dCS gravity, its spin-induced scalar dipole moment yields a scalar flux at $O(\alpha^2\eta^0)$, while in EdGB gravity the scalar monopole charge yields a dipolar scalar flux at $O(\alpha^2\eta^{-2})$. These scalar channels therefore dominate the formal EMRI expansion.

\begin{figure}[tb]
    \centering
    \includegraphics[width=\columnwidth]{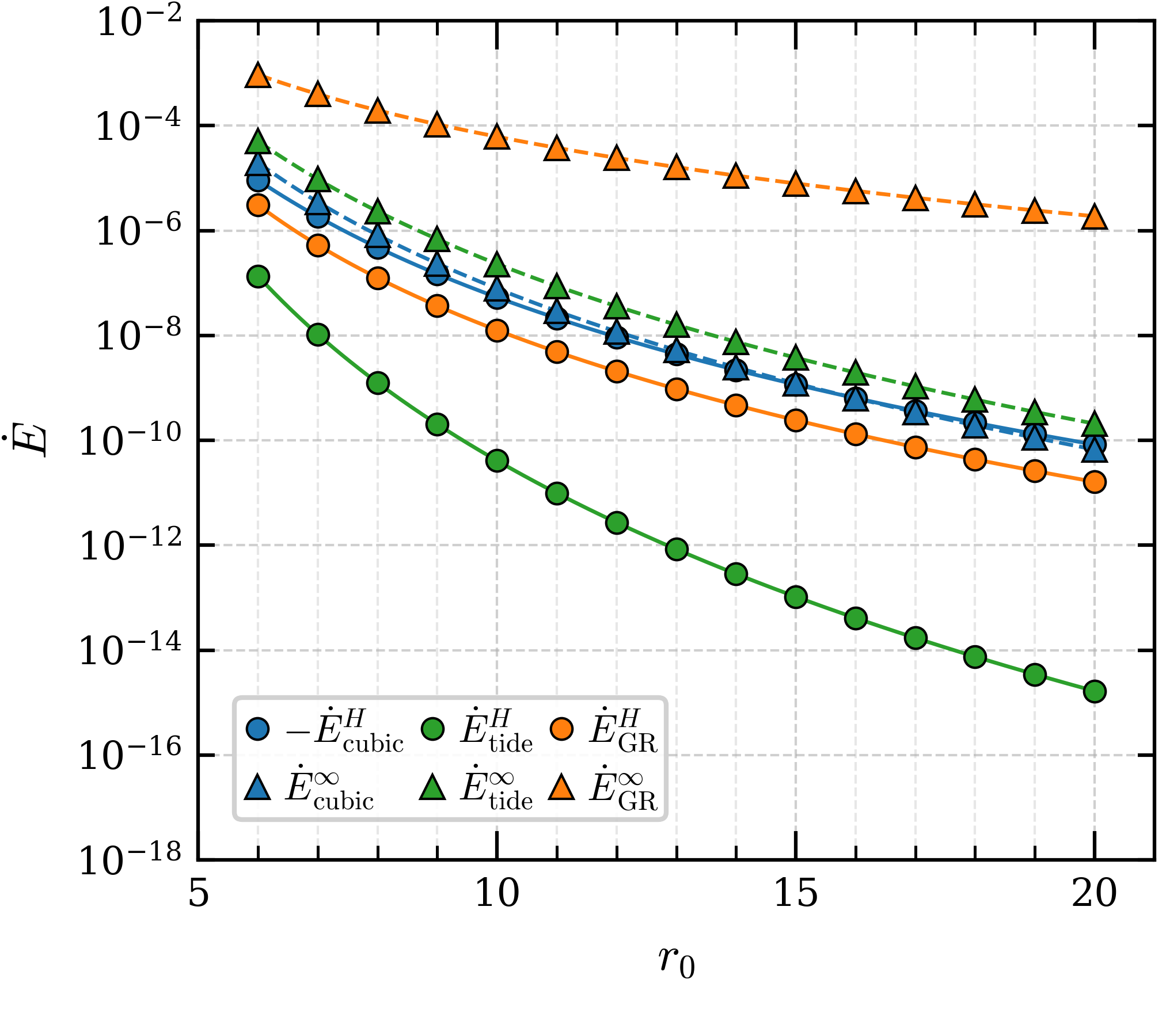}
    \caption{GR energy fluxes and leading beyond-GR corrections through the black-hole horizon ($H$) and at null infinity ($\infty$) as functions of the orbital radius $r_0$. Orange, blue, and green curves denote the GR fluxes, direct cubic-curvature corrections, and finite-size tidal corrections, respectively; circles (triangles) denote horizon (infinity) contributions. The labels ``cubic'' and ``tide'' denote the direct cubic-curvature and finite-size tidal corrections, respectively. The overall factors $\eta^2$ and $\zeta\eta^2$ are removed from the GR fluxes and beyond-GR corrections, respectively. Since the direct cubic-curvature correction to the horizon flux is negative over the displayed range, $-\dot E_{\rm cubic}^{H}$ is shown on the logarithmic scale.}
\label{fig:energy_flux}
\end{figure}

To compute these $O(\zeta\eta^2)$ flux corrections, we first determine the mixed-order radiative perturbations $\Psi_{0,4}^{(1,1)}$. After projection onto spin-weighted spherical harmonics, the modified Teukolsky equations reduce to inhomogeneous radial equations, which we solve using Green functions. Two technical issues arise. First, the higher-curvature sources contain derivatives of the GR metric perturbation up to sixth order; we evaluate them spectrally by expanding the numerical metric perturbations in a Chebyshev basis and using polynomial identities, avoiding repeated numerical differentiation. Second, the Green-function integrals require regularization near the horizon; we isolate the near-horizon contribution and evaluate it using lower incomplete gamma functions. The same strategy can be extended to rotating backgrounds and to cases with asymptotic divergences at infinity. Further details are given in Ref.~\cite{Yang:2026erj}.

With the mixed-order Weyl scalars, the corresponding energy fluxes are extracted from their asymptotic and near-horizon limits. At null infinity, the higher-curvature corrections to the effective gravitational-wave stress-energy tensor fall off faster than $r^{-2}$, so the Isaacson result and the standard relation to $\Psi_4$ remain unchanged~\cite{Stein:2010pn,Isaacson:1968zza}. The leading beyond-GR correction is therefore
\begin{align}
    \frac{d\dot E_\infty^{(1,2)}}{d\Omega} &=  \frac{r^2}{64\pi\omega^2}\left(\Psi_4^{(0,1)}\bar{\Psi}_4^{(1,1)}+\bar{\Psi}_4^{(0,1)}\Psi_4^{(1,1)}\right).
\end{align}
At the horizon, the flux relation is modified by the deformation of the horizon geometry. Generalizing the Hawking-Hartle construction in Refs.~\cite{HawkingHartle1972,Chandrasekhar_1983}, we obtain 
\begin{align}
   \frac{d\dot E_H^{(1,2)}}{d\Omega}&=\frac{M_{\rm BH}}{4\pi\kappa_H}\bigg[2\,\mathrm{Re}\left(\sigma^{(0,1)}\bar{\sigma}^{(1,1)}\right)
    +\rho^{(0,2)}\kappa_H^{(1,0)}+\Phi_{00}^{(1,2)}\bigg].
\end{align}
Here, $\sigma$ and $\rho$ are the shear and expansion of the horizon generators, respectively, while $\kappa_H$ denotes the surface gravity. For the horizon tetrad used here, $\kappa_H=2\epsilon_H$, where $\epsilon_H$ is the Newman-Penrose spin coefficient evaluated on the horizon. The Ricci identities relate these quantities to $\Psi_0^{(0,1)}$ and $\Psi_0^{(1,1)}$, so that the absorbed flux is ultimately determined by $\Psi_0$ together with the local higher-curvature contribution $\Phi_{00}^{(1,2)}$. The calculation summarized above follows Ref.~\cite{Yang:2026erj} for the direct cubic-curvature correction. The tidal worldline terms can be incorporated within the same MTF framework, with the corresponding source construction and flux calculation detailed in the Supplemental Material~\cite{SM}.

The GR fluxes and the two beyond-GR contributions are shown in Fig.~\ref{fig:energy_flux}. The magnitudes of the beyond-GR corrections grow toward smaller orbital radii. Over the range shown, the direct cubic-curvature correction is negative at the horizon and positive at null infinity, while the finite-size tidal corrections are positive. Their growth toward the ISCO reflects the strong-field character of the higher-curvature effects.

\vspace{0.2cm}

\noindent{{\emph{Modified EMRI dynamics}}}---Within the adiabatic approximation, the inspiral is described as a sequence of circular equatorial orbits. For each $(\ell,m)$ mode, $\omega_{\ell m}=m\omega_z$, and hence $\Delta\phi_{\ell m}=m\Delta\phi$, where $\Delta\phi$ is the orbital-phase correction. Since $\omega_z$ determines the observable gravitational-wave frequencies, we compare the GR and modified configurations at fixed $\omega_z$, labeling each frequency by the corresponding GR orbital radius $r_0$ through $\omega_z=\sqrt{M_{\rm BH}/r_0^3}$. The slow inspiral is governed by energy balance,
\begin{align}
    \dot{E}_p=-(\dot{E}_{\infty}+\dot{E}_{H}) .
\end{align}
Here, $E_p$ is the conserved orbital energy of the secondary, while $\dot E_{\infty}$ and $\dot E_H$ are the gravitational-wave energy fluxes at null infinity and through the black-hole horizon, respectively.

Expanding the balance law to linear order in $\zeta$ gives the leading beyond-GR correction to the accumulated phase,
\begin{align}
\begin{split}
    \Delta\phi_{\ell m}= \frac{\zeta}{\eta} m\int \omega_z\Bigg[&\frac{(dE_{p,\rm GR}/dr_0)(\dot E_{\infty,\rm mod}+\dot E_{H,\rm mod})}{(\dot E_{\infty,\rm GR}+\dot E_{H,\rm GR})^2}\nonumber\\
    &-\frac{dE_{p,\rm mod}/dr_0}{\dot E_{\infty,\rm GR}+\dot E_{H,\rm GR}}\Bigg]dr_0 .
\end{split}
\end{align}
Here, the subscripts ``GR'' and ``mod'' denote the GR quantities and the coefficients of their leading beyond-GR corrections, respectively. The first term captures the dissipative correction from the modified energy fluxes, whereas the second encodes the conservative correction to the orbital energy.

We set $\Delta\phi=0$ at the reference frequency $M_{\rm BH}\omega_z=20^{-3/2}$, corresponding to $r_0=20M_{\rm BH}$ in GR, so Fig.~\ref{fig:inspiral_dynamical_phase} shows the accumulated phase correction from this reference frequency toward the innermost stable circular orbit. At the innermost stable circular orbit, the normalized phase shift reaches $\eta\Delta\phi/\zeta\simeq-0.34$. Therefore, a coupling $\zeta\simeq 3\eta$ would produce an accumulated dephasing of order one radian. For a typical EMRI with $\eta\sim10^{-5}$, this corresponds to $\zeta\sim3 \times10^{-5}$, providing an indicative phase-sensitivity scale rather than a parameter-estimation bound.

\begin{figure}[tb]
    \centering
    \includegraphics[width=\columnwidth]{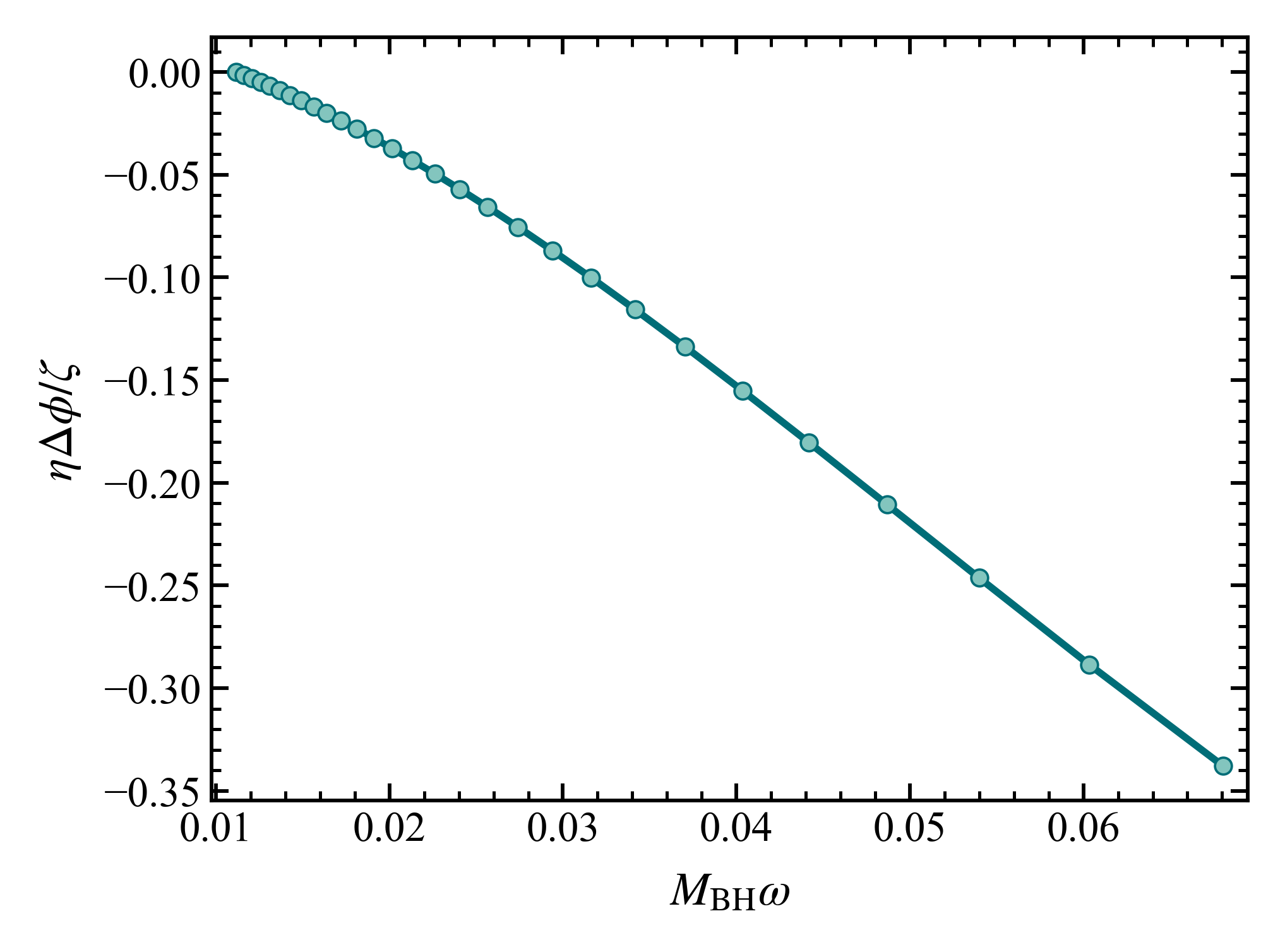}
    \caption{Accumulated modified-gravity correction to the inspiral phase, $\eta\Delta\phi/\zeta$, as a function of the orbital frequency. We set $\Delta\phi=0$ at the reference frequency $M_{\rm BH}\omega_z=20^{-3/2}$, corresponding to $r_0=20M_{\rm BH}$ in GR.}
\label{fig:inspiral_dynamical_phase}
\end{figure}

\vspace{0.2cm}

\noindent{\emph{{EOB extension to comparable masses}}}---EOB waveform models have previously incorporated theory-specific higher-curvature corrections in the remnant quasinormal-mode spectrum \cite{Silva:2022srr}, and a complete EOB inspiral--merger--ringdown construction has been developed for Einstein-scalar-Gauss-Bonnet gravity using post-Newtonian conservative and radiative information \cite{Julie:2024fwy}. Here we construct a complementary model for parity-even cubic gravity in which the strong-field test-mass limit is fixed directly by the modified black hole geometry and the sourced MTF calculation. The former determines the conservative test-mass dynamics, while the latter supplies the gravitational-wave fluxes at null infinity and through the horizon, including both bulk higher-curvature and finite-size tidal contributions.
The model recovers the GR EOB dynamics as $\zeta\rightarrow0$ and the modified test-mass dynamics and the MTF fluxes as $\eta\rightarrow0$.

Hereafter, $M=m_1+m_2$ denotes the total binary mass and $\eta=m_1m_2/M^2$ the symmetric mass ratio; in the test-mass limit, $\eta$ reduces to the mass-ratio parameter used above. For comparable masses, we use $\zeta=\lambda_{\rm ev}l_c^4/M^4$. We construct the conservative dynamics as
\begin{align}
    H_{\mathrm{EOB}}^{\mathrm{MG}}
    =H_{\mathrm{EOB}}^{\mathrm{GR}}
    +\delta H_{\mathrm{EOB}}^{\mathrm{cubic}}
    +\delta H_{\mathrm{EOB}}^{\mathrm{tide}}
    +\delta H_{\mathrm{EOB}}^{\mathrm{PM}},
\label{eq:eob_hamiltonian_schematic}
\end{align}
where the cubic-curvature and tidal terms encode the strong-field test-mass contributions obtained above, while $\delta H_{\mathrm{EOB}}^{\mathrm{PM}}$ incorporates the finite-mass-ratio
conservative information of Ref.~\cite{Wilson-Gerow:2025xhr}. Its overlap with the test-mass contribution is subtracted so that $\delta H_{\mathrm{EOB}}^{\mathrm{PM}}\rightarrow0$ as $\eta\rightarrow0$. The dissipative sector is constructed directly from the fluxes,
\begin{align}
    \mathcal{F}_{E}^{\mathrm{MG}}=\mathcal{F}_{E}^{\mathrm{GR,EOB}}+\delta\mathcal{F}_{E}^{\mathrm{cubic}}+\delta\mathcal{F}_{E}^{\mathrm{tide}} .
\label{eq:eob_flux_schematic}
\end{align}
We retain the beyond-GR flux corrections in their leading test-mass form. Starting from $M\Omega=20^{-3/2}$, the modified strong-field fluxes cover the inspiral interval used here, and no additional post-Newtonian completion is introduced.

The orbital evolution follows from Hamilton's equations,
\begin{align}
    \dot{R}&=\frac{\partial H_{\mathrm{EOB}}^{\mathrm{MG}}}{\partial P_R},&\dot{P}_R&=-\frac{\partial H_{\mathrm{EOB}}^{\mathrm{MG}}}{\partial R}+\mathcal{F}_R,\notag\\
    \dot{\phi}&=\frac{\partial H_{\mathrm{EOB}}^{\mathrm{MG}}}{\partial P_\phi},&\dot{P}_\phi&=\mathcal{F}_\phi .
\label{eq:eob_hamilton_equations}
\end{align}
Here, $(R,\phi)$ are the EOB radial and azimuthal coordinates and $(P_R,P_\phi)$ their conjugate momenta. For the quasi-circular evolution considered here, we adopt $\mathcal{F}_R=0$ and $\mathcal{F}_\phi=-\mathcal{F}^{\mathrm{MG}}_E/\Omega$, with $\Omega=\dot{\phi}$. We implement these modifications in TEOBResumS~\cite{Nagar:2018zoe,Nagar:2019wds} to construct complete inspiral-merger-ringdown waveforms. Following the strategy of Ref.~\cite{Julie:2024fwy}, we retain the GR-calibrated merger and postpeak baseline in the absence of theory-specific numerical-relativity calibration, while incorporating the coupling-dependent remnant properties and cubic-gravity quasinormal modes~\cite{Cano:2023jbk}; technical details are given in the Supplemental Material~\cite{SM}.

The resulting equal-mass waveforms are shown in Fig.~\ref{fig:imr_waveforms}. All evolutions start at the common reference frequency $M\Omega=20^{-3/2}$ with $t=0$. In Fig.~\ref{fig:imr_waveforms}(a), all waveforms are shown relative to the GR waveform-amplitude peak, $(t-t_{A_{22},{\rm peak}}^{\rm GR})/M$, while retaining their common initial time and phase reference. The modified waveforms remain close to GR for small couplings, while a visible dephasing develops before merger for $\zeta\gtrsim10^{-2}$. At $\zeta=10^{-2}$, the $(2,2)$-mode phase difference reaches $|\Delta\phi_{22}|\simeq0.60\,\mathrm{rad}$ during the inspiral-plunge evolution, and its waveform-amplitude peak occurs $1.83M$ earlier than in GR. In Fig.~\ref{fig:imr_waveforms}(b), each waveform is aligned in time and phase at its own waveform-amplitude peak, while its amplitude is left unchanged. The remaining differences therefore characterize coupling-dependent changes in the near-merger and ringdown morphology. A joint ringdown analysis constrains the parity-even cubic-gravity length scale to $l_c\lesssim34.3\,\mathrm{km}$ at $95\%$ credibility \cite{Maenaut:2024oci}. For a representative equal-mass binary with $m_1=m_2=30M_\odot$, this corresponds to $\zeta\lesssim2\times10^{-2}$ for $\lambda_{\rm ev}=1$. Larger values of $\zeta$ are included only to emphasize the waveform modifications.

\vspace{0.2cm}


\noindent{{\emph{Discussion}}}---The EOB construction used here extrapolates the MTF results toward comparable masses. A complementary route is to extend the waveform systematically within the small-mass-ratio expansion. Recent multiscale self-force frameworks connect the post-adiabatic inspiral to transition, plunge, and merger-ringdown directly within black hole perturbation theory~\cite{Miller:2020bft,Kuchler:2024esj,Kuchler:2025hwx,Roy:2025kra}. Applied to the mixed-order sources derived here, they could provide an independent waveform construction for asymmetric binaries. These approaches therefore complement, rather than replace, the EOB extrapolation toward comparable masses.

Extensions to theories with additional dynamical fields require further ingredients. For example, in dynamical Chern-Simons gravity, the radiative Weyl scalars must be evolved together with the pseudoscalar perturbation, forming a coupled tensor-scalar system~\cite{Wagle:2023fwl,Li:2025fci}. Effective scalar-charge treatments capture important scalar-radiation effects in the EMRI regime~\cite{Maselli:2020zgv}, but by themselves do not supply the coupled tensor-scalar dynamics or finite-mass-ratio information needed for a complete comparable-mass waveform.

The MTF calculation presented here provides controlled strong-field information in the test-mass limit, while its EOB extension extrapolates this information to comparable masses. Results from test-mass and self-force calculations have shown remarkable accuracy beyond their nominal regime in GR~\cite{LeTiec:2011bk,Rifat:2019ltp,vandeMeent:2020xgc, PhysRevLett.130.241402}, but the corresponding extrapolation in the cubic gravity considered here, or beyond-GR theories in general, has not yet been benchmarked at comparable mass ratios. Future self-force calculations and numerical relativity simulations can directly test and refine this extrapolation.

\acknowledgments
\vspace{0.2cm}
\emph{Acknowledgments}---We are grateful to Pablo Cano and Luis Lehner for insightful discussions. This work makes use of the Black Hole Perturbation Toolkit. H.~Y. is supported by the Natural Science Foundation of China (Grant 12573048). N.~K. is supported by the Shuimu fellowship of Tsinghua University. D.~L. acknowledges support from the Simons Foundation (via Award No. 896696), the Simons Foundation International (via Grant No. SFI-MPS-BH-00012593-01), and the NSF (via Grants No.~PHY-2512423). OpenAI Codex was used to assist with code development and debugging, numerical consistency checks, figure preparation, and manuscript editing. All AI-assisted outputs have been independently verified by the authors.

\bibliography{reference_new}
\end{document}


\setcounter{secnumdepth}{2}

\title{Supplemental Material for\\
``From the Test-Mass Limit to Binary Black-Hole Waveforms \\
in Higher-Derivative Gravity''}

\author{Chaoyi Yang}
\affiliation{Department of Astronomy, Tsinghua University, Beijing 100084, China}

\author{Neev Khera \orcidlink{0000-0003-3515-2859}}
\affiliation{Department of Astronomy, Tsinghua University, Beijing 100084, China}

\author{Dongjun Li \orcidlink{0000-0002-1962-680X}}
\affiliation{\UIUC}

\author{Huan Yang \orcidlink{0000-0002-9965-3030}}
\affiliation{Department of Astronomy, Tsinghua University, Beijing 100084, China}

\maketitle

\section{Tidal Correction to the Energy Flux from Love Numbers}\label{sec:SM_tidal_flux}

Here, we provide the details of the tidal-flux calculation summarized in the Letter. The leading quadrupolar response of the secondary is described by the worldline action
\begin{align}
    S_{\mathrm{pp}}&=-m_p\int d\tau+\frac{c_E}{2}\int d\tau\,E_{\mu\nu}E^{\mu\nu}+\frac{c_B}{2}\int d\tau\,B_{\mu\nu}B^{\mu\nu}.
\end{align}
Here, $E_{\mu\nu}$ and $B_{\mu\nu}$ are the electric- and magnetic-type tidal tensors defined in the Letter, and $C_{\mu\nu\rho\sigma}$ is the Weyl tensor. Using
\begin{align}
    E_{\mu\nu}E^{\mu\nu}-B_{\mu\nu}B^{\mu\nu}&=\frac{1}{8}\,C_{\mu\nu\rho\sigma}C^{\mu\nu\rho\sigma},
\end{align}
we rewrite the tidal action as
\begin{align}
    S_{\mathrm{pp}}&=-m_p\int d\tau+\frac{c_E+c_B}{2}\int d\tau\,E_{\mu\nu}E^{\mu\nu}-\frac{c_B}{16}\int d\tau\,C_{\mu\nu\rho\sigma}C^{\mu\nu\rho\sigma}.
\label{eq:action_tidal}
\end{align}

For the parity-preserving cubic interaction considered here, Ref.~\cite{Cano:2026hlv} gives the quadrupolar electric- and magnetic-type Love numbers $k_2^+=28\lambda_{\rm ev}l_c^4/m_p^4$ and $k_2^-=-20\lambda_{\rm ev}l_c^4/m_p^4$, respectively. Eqs.~(A12) and (A13) of Ref.~\cite{Wang:2026qst} relate the quadrupolar Love numbers obtained from the static black-hole response to the coefficients of the associated electric- and magnetic-type worldline operators. Relative to the normalization used in Ref.~\cite{Wang:2026qst}, our worldline action contains an additional factor of $1/2$ multiplying each tidal operator. Accordingly, in our convention these relations become $c_E=m_p^5k_2^+/3$ and $c_B=m_p^5k_2^-/3$. Substituting the Love numbers above then gives $c_E=28m_p\lambda_{\rm ev}l_c^4/3$ and $c_B=-20m_p\lambda_{\rm ev}l_c^4/3$.

The Weyl-squared term in Eq.~\eqref{eq:action_tidal} reduces to the Kretschmann-scalar interaction on the Ricci-flat GR background and has the same operator structure as the worldline term generated in the field-redefined $RK$ theory studied in Appendix D of Ref.~\cite{Yang:2026erj}. Its contribution can therefore be obtained from the corresponding $RK$ results by an overall rescaling of the coupling. With the conventions used here, the required factor relative to the $RK$ normalization is $-c_B/(8m_p\lambda_{\rm ev}l_c^4)=5/6$. This rescaling applies to the associated stress-energy tensor, orbital correction, and modified Teukolsky source. The electric-type interaction, which depends explicitly on the particle four-velocity, is evaluated separately below.

Since both Wilson coefficients are proportional to $m_p\lambda_{\rm ev}l_c^4$, the tidal interactions first modify the radiative perturbation at $O(\zeta\eta)$ and do not alter the stationary background at $O(\zeta^1,\eta^0)$. At mixed order, the source receives two types of contributions: the explicit stress-energy tensor generated by the tidal worldline operators and the correction obtained by evaluating the leading $O(\eta)$ point-particle source on the $O(\zeta)$-shifted circular orbit. Below we derive these contributions for the electric-type operator; the Weyl-squared contribution is obtained directly from the field-redefined $RK$ result described above.

\subsection{Electric-type tidal contribution}
\label{sec:SM_electric_tidal}
To avoid ambiguity with the original electric-type Wilson coefficient, we define $\tilde{c}_{E}\equiv c_E+c_B$. The $E^2$ interaction in Eq.~\eqref{eq:action_tidal} can then be written as $S_{E}=(\tilde{c}_{E}/2)\int d\tau\,E_{\mu\nu}E^{\mu\nu}$. After the decomposition in Eq.~\eqref{eq:action_tidal}, no independent $B_{\mu\nu}B^{\mu\nu}$ operator remains to be varied separately: its contribution is split between the $E^2$ interaction through $\tilde{c}_{E}=c_E+c_B$ and the Weyl-squared interaction proportional to $c_B$. We treat the former explicitly below, while the latter is inherited from the field-redefined $RK$ calculation described above and is therefore not rederived here. Defining $\mathcal I_E\equiv E_{\mu\nu}E^{\mu\nu}$, the variation of the worldline interaction with respect to the metric can be written as
\begin{align}
    \delta S_{E}=\frac{\tilde{c}_{E}}{2}\int d\tau\Bigg[\left(-2E^{\mu\rho}E^{\nu}{}_{\rho}+\frac{3}{2}\mathcal I_E u^\mu u^\nu\right)\delta g_{\mu\nu}+2E^{\mu\nu}u^\alpha u^\beta\delta C_{\mu\alpha\nu\beta}\Bigg].
\end{align}
After integrating by parts to remove the derivatives acting on the metric variation through $\delta C_{\mu\alpha\nu\beta}$, this expression determines the electric-type contribution to the particle stress-energy tensor and hence to the Newman--Penrose source.

The tidal interaction also modifies the particle motion. Following the worldline variational procedure of Appendix D of Ref.~\cite{Yang:2026erj}, we introduce an arbitrary parameter $\lambda$, with $d\tau=\mathcal N d\lambda$, $\mathcal N=\sqrt{-g_{\mu\nu}\dot{x}^\mu\dot{x}^\nu}$, and $u^\mu=\dot{x}^\mu/\mathcal N$. The corresponding worldline Lagrangian is
\begin{align}
    \mathcal{L}=\left(-m_p+\frac{\tilde{c}_{E}}{2}\mathcal I_E\right)\mathcal N .
\end{align}
Defining $A_\mu\equiv\partial\mathcal I_E/\partial u^\mu$ and using $u^\mu A_\mu=4\mathcal I_E$, the derivatives entering the Euler--Lagrange equations, after setting $\lambda=\tau$, are 
\begin{align}
    \frac{\partial\mathcal L}{\partial\dot{x}^\mu} &= m_pu_\mu+\frac{\tilde{c}_{E}}{2}\left(A_\mu+3\mathcal I_Eu_\mu\right),\\
    \frac{\partial\mathcal L}{\partial x^\mu}&=\frac{m_p}{2}\partial_\mu g_{\alpha\beta}u^\alpha u^\beta+\frac{\tilde{c}_{E}}{2}(\partial_\mu\mathcal I_E)_u+\frac{3\tilde{c}_{E}}{4}\mathcal I_E\partial_\mu g_{\alpha\beta}u^\alpha u^\beta .
\end{align}
Here, $(\partial_\mu\mathcal I_E)_u$ denotes differentiation with respect to $x^\mu$ at fixed components of $u^\alpha$. To linear order in $\tilde{c}_{E}$, the Euler--Lagrange equations give the four-acceleration $a_\mu\equiv u^\nu\nabla_\nu u_\mu$ as
\begin{align}
    a_\mu = \frac{\tilde{c}_{E}}{2m_p} \left[ (\partial_\mu\mathcal I_E)_u -\frac{dA_\mu}{d\tau} -3u_\mu\frac{d\mathcal I_E}{d\tau} \right].
\label{eq:E2_EOM}
\end{align}

For a circular equatorial orbit, we solve Eq.~\eqref{eq:E2_EOM} together with $u^\mu u_\mu=-1$ while keeping the orbital frequency $\omega_z$ fixed. Writing $r=r_0+\delta r_0$ and $u^t\rightarrow u^t+\delta u^t$, we find $\delta u^t=0$ and $\delta r_0=
6\tilde{c}_{E}M_{\rm BH}(2M_{\rm BH}-r_0)^2(u^t)^2/(m_pr_0^6).$ Thus, the tidal interaction does not modify the stationary background at
$O(\zeta^1,\eta^0)$, but it shifts the particle trajectory at $O(\zeta)$.

The complete mixed-order source is obtained by combining the explicit stress-energy contribution from $S_{E}$ with the correction generated by the shifted orbit in the leading point-particle source. Projecting the result onto the Newman-Penrose tetrad determines the mixed-order Newman-Penrose Ricci scalars $\Phi_{ab}^{(1,1)}$, which we verify to satisfy the corresponding Bianchi identities. Since the tidal sector has no stationary $O(\zeta,\eta^0)$ contribution, the corresponding modified Teukolsky equations contain only the mixed-order particle source,
\begin{align}
    H_{0}^{\rm GR}\Psi_{0}^{(1,1)}&= \mathcal{S}^{(1,1)},\\
    H_{4}^{\rm GR}\Psi_{4}^{(1,1)}&= \mathcal{T}^{(1,1)}.
\label{eq:E2_MTE}
\end{align}
Here, $H_{0}^{\rm GR}$ and $H_{4}^{\rm GR}$ denote the standard GR Teukolsky operators. The source terms are constructed entirely from the $\Phi_{ab}^{(1,1)}$ above, and we solve the equations using the same frequency-domain Green-function procedure as in Ref.~\cite{Yang:2026erj}.

At null infinity, the leading tidal correction is obtained from the interference between the GR and mixed-order $\Psi_4$ perturbations,
\begin{align}
    \frac{d\dot E_\infty^{(1,2)}}{d\Omega}=\frac{r^2}{64\pi\omega^2}\left(\Psi_4^{(0,1)}\bar{\Psi}_4^{(1,1)}+\bar{\Psi}_4^{(0,1)}\Psi_4^{(1,1)}\right).
\label{eq:infinityflux}
\end{align}
Here, $\omega$ denotes the Fourier frequency of the corresponding $(\ell,m)$ mode. Integrating over the sphere and summing over the radiative modes gives the total tidal correction to the flux at null infinity.

The horizon calculation is simpler than that for the direct cubic-curvature sector. The mixed-order Ricci scalars generated by the tidal worldline interaction are supported only on the particle worldline and therefore vanish on the horizon, while the tidal sector introduces no $O(\zeta,\eta^0)$ background deformation. The relevant Newman--Penrose relations therefore reduce to
\begin{align}
    \rho^{(0,2)}&=-\frac{1}{2\epsilon_H} \left|\sigma^{(0,1)}\right|^2,\\
    \sigma^{(0,1)}&=-\frac{1}{i\omega+2\epsilon_H}\Psi_0^{(0,1)},\\
    \sigma^{(1,1)}&=-\frac{1}{i\omega+2\epsilon_H}\Psi_0^{(1,1)},
\label{eq:spin_coeff_horizon}
\end{align}
Here, $\rho$ and $\sigma$ denote the expansion and shear of the horizon generators, respectively, and $\epsilon_H$ is the background Newman--Penrose spin coefficient evaluated on the horizon. The leading tidal correction to the absorbed flux is then
\begin{align}
    \frac{d\dot E_H^{(1,2)}}{d\Omega}=\frac{M_{\rm BH}}{8\pi\epsilon_H}\left(\sigma^{(0,1)}\bar{\sigma}^{(1,1)}+\bar{\sigma}^{(0,1)}\sigma^{(1,1)}\right).
\label{eq:horizonflux}
\end{align}

Finally, combining the $E^2$ contribution weighted by $\tilde{c}_{E}=c_E+c_B$ with the Weyl-squared contribution inherited from the field-redefined $RK$ calculation yields the complete $O(\zeta\eta^2)$ tidal corrections to the horizon and infinity fluxes used in the Letter.


\section{Orbital Energy and Accumulated Phase} \label{sec:SM_phase}

Using the adiabatic evolution and fixed-$\omega_z$ prescription described in the Letter, we expand the orbital energy and gravitational-wave fluxes as
\begin{align}
    E_p 
    &= \eta E_{p,\rm GR} +\eta \zeta E_{p,\rm mod}+\mathcal{O}(\eta^2 \zeta)\,, \nonumber \\
    \dot{E}_{\infty} 
    &=\eta^2\dot{E}_{\infty,\rm GR}+\eta^2\zeta\dot{E}_{\infty,\rm mod}
    +\mathcal{O}(\eta^3 \zeta)\,, \nonumber \\
    \dot{E}_{H} 
    &=\eta^2\dot{E}_{H,\rm GR}+\eta^2\zeta\dot{E}_{H,\rm mod}
    +\mathcal{O}(\eta^3\zeta)\,,\label{eq:SM_phase_expansion}
\end{align}
Substituting Eq.~\eqref{eq:SM_phase_expansion} into the balance law $\dot E_p=-(\dot E_\infty+\dot E_H)$, the physical phase correction may be written as $\Delta\phi_{\ell m}=(\zeta/\eta)\Delta\Phi_{\ell m}$, where
\begin{align}
\begin{split}
    \Delta\Phi_{\ell m}= m\int \omega_z\Bigg[\frac{(dE_{p,\rm GR}/dr_0)(\dot E_{\infty,\rm mod}+\dot E_{H,\rm mod})}{(\dot E_{\infty,\rm GR}+\dot E_{H,\rm GR})^2}-\frac{dE_{p,\rm mod}/dr_0}{\dot E_{\infty,\rm GR}+\dot E_{H,\rm GR}}\Bigg]dr_0 .
\end{split}
\label{eq:SM_phase_correction}
\end{align}
The first term inside the brackets in Eq.~\eqref{eq:SM_phase_correction} is the dissipative contribution from the modified energy fluxes, while the second is the conservative contribution from the modified orbital energy. Since $\Delta\phi_{\ell m}=m\Delta\phi$, with $\Delta\phi$ the orbital-phase correction, the quantity plotted in the Letter is $\eta\Delta\phi/\zeta$.

For the stationary Killing vector $k^\mu=(1,0,0,0)$, the conserved particle energy is $E_p=-k^\mu p_\mu$, where the canonical momentum is obtained from the worldline Lagrangian as $p_\mu=\partial\mathcal{L}/\partial\dot{x}^\mu$ after setting $\lambda=\tau$. At $O(\zeta)$, we decompose the modified contribution as
\begin{align}
    E_{p,\rm mod}&=-k^\mu\left(p_\mu^{\rm cubic}+p_\mu^{\rm tide}\right),\\
    p_\mu^{\rm cubic}&=m_p\,\delta u_\mu^{\rm cubic},\\
    p_\mu^{\rm tide}&=m_p\,\delta u_\mu^{\rm tide}+\frac{c_E+c_B}{2}\left(A_\mu+3\mathcal I_Eu_\mu\right)+\frac{c_B}{16}C_{\alpha\beta\gamma\delta}C^{\alpha\beta\gamma\delta}u_\mu .
\label{eq:SM_particle_momentum}
\end{align}
For the direct cubic-curvature contribution, $\delta u^\mu_{\rm cubic}$ is obtained from the modified circular orbit at fixed $\omega_z$ following Ref.~\cite{Yang:2026erj}. For the tidal contribution, the corresponding equations of motion, together with the field-redefined $RK$ result for the Weyl-squared term, give $\delta u^t_{\rm tide}=0$, while the orbital radius is shifted at $O(\zeta)$. Thus, $\delta u_\mu^{\rm tide}$ accounts for the change induced by the shifted orbit, whereas the remaining terms in $p_\mu^{\rm tide}$ are the explicit canonical-momentum corrections generated by the tidal worldline operators. The quantities $\mathcal I_E$ and $A_\mu$ are defined in Sec.~\ref{sec:SM_electric_tidal}. 

We fix the integration constant by setting $\Delta\phi=0$ at the reference frequency $M_{\rm BH}\omega_z=20^{-3/2}$, corresponding to $r_0=20M_{\rm BH}$ in GR, so that the phase reported in the Letter is accumulated from this reference frequency toward the ISCO.

\section{Inspiral-Merger-Ringdown Waveform in the EOB Framework}\label{sec:SM_EOB}

Here, we provide the technical details of the EOB construction summarized in the Letter. We use the same notation,
\begin{align}
    M=m_1+m_2,\qquad  \mu=\frac{m_1m_2}{M}=\eta M,\qquad \eta=\frac{m_1m_2}{M^2},
\end{align}
and introduce $u=M/R$, $p_R=P_R/\mu$, and $p_\phi=P_\phi/(\mu M)$.

\subsection{Conservative dynamics} \label{sec:SM_EOB_conservative}
The EOB framework maps the conservative two-body dynamics onto the motion of an effective particle of mass $\mu$ in an effective geometry, supplemented by nongeodesic interactions. For the nonspinning system considered here, we parameterize the effective metric as
\begin{align}
    ds_{\rm eff}^2=-A(R;\eta,\zeta)dt^2+\frac{dR^2}{B(R;\eta,\zeta)}+R^2d\Omega^2 .
\end{align}
Here, $R$ is the EOB area-radius coordinate, while $A$ and $B$ are the two metric potentials. In the test-mass limit, the effective geometry is required to reproduce the nonspinning cubic-gravity black-hole solution of Ref.~\cite{Yang:2026erj},
\begin{align}
    ds^2=-(1-\frac{2M}{r})\left[1-\zeta H_1(r)\right]dt^2+(1-\frac{2M}{r})^{-1}\left[1+\zeta H_3(r)\right]dr^2+r^2\left[1+\zeta H_3(r)\right]d\Omega^2,
\end{align}
Here, $H_1(r)$ and $H_3(r)$ are the radial correction functions given in Ref.~\cite{Yang:2026erj}. Comparing the angular parts of the two metrics gives $R=r[1+\zeta H_3(r)/2]+\mathcal O(\zeta^2)$. Rewriting the remaining metric components in terms of $R$ and defining $u=M/R$ then gives the test-mass metric potentials
\begin{align}
    A(u;0,\zeta)&=1-2u+40\zeta u^7,\\
    B(u;0,\zeta)&=1-2u+\zeta\left(216u^6-392u^7\right).
\end{align}

Given this effective geometry, the conservative dynamics of the effective particle is formulated in terms of its canonical four-momentum $P_\mu$, with the effective Hamiltonian defined by the conserved energy $H_{\rm eff}\equiv-P_t$. The geodesic sector is determined by the mass-shell condition $g_{\rm eff}^{\mu\nu}P_\mu P_\nu=-\mu^2$, which fixes how the metric potentials $A$ and $B$ enter $H_{\rm eff}$. Beyond this geodesic sector, the EOB Hamiltonian contains an additional nongeodesic potential $Q$, which is not determined by the effective metric. In the GR baseline, $Q_{\rm GR}^{\rm EOB}$ encodes finite-mass-ratio contributions beyond the geodesic sector. In the EOB gauge adopted here, it is supplemented by the tidal finite-size interaction and the finite-mass-ratio cubic PM correction. Using the dimensionless momenta introduced above, the full effective Hamiltonian is
\begin{align}
    \widehat{H}_{\mathrm{eff}}^{2}=A(u;\eta,\zeta)\left[1+p_\phi^2u^2+B(u;\eta,\zeta)p_R^2+Q(u,p_R,p_\phi;\eta,\zeta)\right],
\label{eq:effective_hamiltonian}
\end{align}
where $\widehat{H}_{\mathrm{eff}}\equiv H_{\mathrm{eff}}/\mu$. The effective energy is related to the physical two-body Hamiltonian through the standard EOB energy map,
\begin{align}
    H_{\mathrm{EOB}}=M\sqrt{1+2\eta\left(\widehat{H}_{\mathrm{eff}}-1\right)}.
\label{eq:eob_mapping}
\end{align}

Away from the test-mass limit, we keep the test-mass cubic corrections in $A$ and $B$ while replacing their GR parts by the standard finite-mass-ratio EOB potentials,
\begin{align}
    A(u;\eta,\zeta)&=A_{\mathrm{GR}}^{\mathrm{EOB}}(u;\eta)+40\zeta u^{7},\\
    B(u;\eta,\zeta)&=B_{\mathrm{GR}}^{\mathrm{EOB}}(u;\eta)+\zeta\left(216u^{6}-392u^{7}\right).
\label{eq:eob_modified_potentials}
\end{align}
By construction, these potentials reproduce the cubic-gravity test-particle geodesic dynamics as $\eta\rightarrow0$ and reduce to the standard GR EOB potentials as $\zeta\rightarrow0$. The additional finite-mass-ratio cubic information is incorporated below through $Q_{\rm cubic}^{\rm PM}$.

We next specify the nongeodesic potential $Q$, which we decompose as
\begin{align}
    Q=Q_{\mathrm{GR}}^{\mathrm{EOB}}+Q_{\rm tide}+Q_{\rm cubic}^{\rm PM}.
\end{align}
Here, $Q_{\rm GR}^{\rm EOB}$ is inherited unchanged from the TEOBResumS GR baseline. For the Giotto model adopted here, the GR non-geodesic potential is given at the third post-Newtonian order by
\begin{equation}
    Q_{\rm GR}^{\rm EOB}(u,p_{R_*};\eta)
    =2\eta(4-3\eta)u^2p_{R_*}^4\,,
\end{equation}
where $p_{R_*}\equiv\sqrt{A B}\,p_R$ is the momentum conjugate to the tortoise radial coordinate $R_*$, defined by $dR_*/dR=(A B)^{-1/2}$. This contribution vanishes for circular motion, $p_{R_*}=0$, as well as in the test-mass limit, $\eta\rightarrow0$. It therefore represents a finite-mass-ratio correction associated with noncircular motion and becomes relevant during the plunge. The remaining two terms encode the tidal finite-size interaction and the finite-mass-ratio cubic PM correction, respectively.

We first consider the tidal contribution, which follows from the worldline action introduced in Sec.~\ref{sec:SM_tidal_flux}. To incorporate this interaction into the EOB Hamiltonian, we evaluate the tidal invariants on the Schwarzschild background and rewrite them in terms of the EOB variables $u$ and $p_\phi$. For the circular equatorial motion considered here, this is achieved using the zeroth-order four-velocity normalization together with $P_\phi=\mu u_\phi$. Since we retain only terms linear in the Wilson coefficients, the corresponding zeroth-order GR relations are sufficient. The resulting tidal correction to the effective Hamiltonian is represented by the nongeodesic potential $Q_{\rm tide}$. Explicitly, the tidal invariants take the form
\begin{align}
    E_{\mu\nu}E^{\mu\nu}&=\frac{6u^6}{M^4}\left(1+3p_\phi^2u^2+3p_\phi^4u^4\right),\\
    B_{\mu\nu}B^{\mu\nu} &=\frac{18u^6}{M^4}\left(p_\phi^2u^2+p_\phi^4u^4\right).
\end{align}
Substituting these expressions into the worldline action and performing the Legendre transformation gives the tidal contribution to the reduced effective Hamiltonian. For a binary with $q\equiv m_1/m_2\geq1$, we take $m_2$ to be the tidally responding black hole. The Wilson coefficients obtained in Sec.~\ref{sec:SM_tidal_flux} are proportional to $m_2$, whereas the reduced effective Hamiltonian is normalized by $\mu$. The corresponding reduced tidal interaction therefore carries the finite-mass-ratio factor $m_2/\mu=(1+q)/q$. Using the Wilson coefficients obtained above, we find
\begin{align}
    Q_{\rm tide}(u,p_\phi)=-\zeta\frac{1+q}{q} \left(56u^6+48p_\phi^2u^8+48p_\phi^4u^{10}\right).
\label{eq:eob_tidal_Q}
\end{align}
This factor follows from the reduced-Hamiltonian normalization and approaches unity in the test-mass limit. Although the tidal invariants are evaluated for circular equatorial motion, we do not further eliminate $p_\phi$ using the circular-orbit relation $p_\phi=p_\phi^{\rm circ}(u)$. We use the resulting momentum-dependent form of $Q_{\rm tide}$ in the subsequent EOB evolution.

We finally incorporate the finite-mass-ratio conservative information from Ref.~\cite{Wilson-Gerow:2025xhr}. We specialize the result of Ref.~\cite{Wilson-Gerow:2025xhr} to the $c_{6,1}$ sector in the field-redefinition basis $c_{6,2}=0$, which corresponds to the parity-even cubic interaction studied here, with $c_{6,1}(l_c/M)^4$ identified with our coupling $\zeta$. Reference \cite{Wilson-Gerow:2025xhr} computes the gauge-invariant radial action $i_r(\gamma,b)$ through 3PM, where $\gamma$ is the relative boost factor, identified through the EOB energy map with the dimensionless effective energy $\widehat H_{\rm eff}$, and $b$ is the impact parameter. The corresponding dimensionless real EOB energy is $\Gamma\equiv H_{\rm EOB}/M=\sqrt{1+2\eta(\gamma-1)}$. The $c_{6,1}$ contributions relevant here enter the 2PM and 3PM radial actions given in Eqs.~(11) and (12) of \cite{Wilson-Gerow:2025xhr},
\begin{align}
    i_r^{\rm 2PM}&=\frac{\pi G^2m_1^2m_2}{b}\frac{h_1(\gamma)}{\sqrt{\gamma^2-1}}+(1\leftrightarrow2),\\
    i_r^{\rm 3PM}&=\frac{G^3m_1^3m_2}{b^2}\frac{h_2(\gamma)}{(\gamma^2-1)^{5/2}}+(1\leftrightarrow2)+\frac{G^3m_1^2m_2^2}{b^2}\left[\frac{h_3(\gamma)}{(\gamma^2-1)^{5/2}}+\frac{h_4(\gamma)\operatorname{arccosh}\gamma}{(\gamma^2-1)^3}
\right],
\end{align}
The functions $h_i$ are the coupling-dependent velocity polynomials given explicitly in Appendix A of Ref.~\cite{Wilson-Gerow:2025xhr}. In the matching below, we retain only their $c_{6,1}$ contributions.

To isolate the cubic terms relevant for the matching, we write the radial-momentum equation schematically as
\begin{align}
    p_R^2=\gamma^2-1-p_\phi^2u^2+W_6(\gamma,\eta)u^6+W_7(\gamma,\eta)u^7+\cdots .
\label{eq:PM_radial_equation}
\end{align}
The coefficients $W_6$ and $W_7$ are determined by inserting Eq.~\eqref{eq:PM_radial_equation} into the reduced radial action $i_r/\mu=2\int p_R\,dR$ and matching it to the corresponding reduced $c_{6,1}$ contributions in Eqs.~(11) and (12) of Ref.~\cite{Wilson-Gerow:2025xhr}. This gives $ W_6^{\rm WG}(\gamma,\eta)=-36\zeta \Gamma^{-5}(\gamma^2-1),W_7^{\rm WG}(\gamma,\eta)=\zeta\Gamma^{-6}B_7(\gamma,\eta)$, where the function $B_7(\gamma,\eta)$ is
\begin{align}
\begin{split}
    B_7(\gamma,\eta)
    =-\frac{256}{7}(1-2\eta)(8\gamma^2-1)+480\eta\,\frac{\gamma\left(2\gamma^6-7\gamma^4+2\gamma^2-42\right)}{(\gamma^2-1)^2}+4320\eta\,\frac{(4\gamma^2+1)\operatorname{arccosh}\gamma}{(\gamma^2-1)^{5/2}}.
\end{split}
\end{align}
For bound configurations, the last term is defined by the analytic continuation $\operatorname{arccosh}\gamma/(\gamma^2-1)^{5/2}\rightarrow\arccos\gamma/(1-\gamma^2)^{5/2}$. Although the individual terms in $B_7(\gamma,\eta)$ appear singular as $\gamma\rightarrow1$, their combination has a finite limit.

The coefficients $W_6^{\rm WG}$ and $W_7^{\rm WG}$ obtained from the full radial action contain both probe-limit and finite-mass-ratio information. Taking $\eta\rightarrow0$ gives $ W_6^{(0)}(\gamma)=-36\zeta(\gamma^2-1), W_7^{(0)}(\gamma)=-256\zeta(8\gamma^2-1)/7$. These terms reproduce the probe-limit cubic contribution already encoded in the cubic-gravity black-hole geometry through the potentials $A$ and $B$. To avoid double counting, we subtract this overlap and define the finite-mass-ratio remainder $\Delta W_i(\gamma,\eta)\equiv W_i^{\rm WG}(\gamma,\eta)-W_i^{(0)}(\gamma)$. Since the PM correction is already linear in $\zeta$, its energy arguments can be evaluated on the EOB dynamics at zeroth order in $\zeta$. We therefore define $\gamma_0\equiv\widehat H_{\rm eff}^{\rm GR}$ and $\Gamma_0\equiv H_{\rm EOB}^{\rm GR}/M=\sqrt{1+2\eta(\gamma_0-1)}$. The coefficients entering our Hamiltonian are therefore
\begin{align}
    \Delta W_6&=-36\zeta(\gamma_0^2-1)\left(\Gamma_0^{-5}-1\right),\\
    \Delta W_7&=\zeta\Gamma_0^{-6}B_7(\gamma_0,\eta)+\frac{256}{7}\zeta(8\gamma_0^2-1).
\end{align}
The remaining step is to convert this radial-momentum correction into the nongeodesic EOB potential $Q_{\rm cubic}^{\rm PM}$. At the orders retained here, the momentum dependence of $Q_{\rm GR}^{\rm EOB}$ does not affect this conversion, since its leading contribution starts at $O(u^2p_R^4)$ and its variation induced by $\delta(p_R^2)=O(\zeta u^6)$ therefore enters only at $O(\zeta u^8)$. Thus, at fixed $\gamma$ and $p_\phi$, Eq.~\eqref{eq:effective_hamiltonian} gives $B\,\delta(p_R^2)+Q_{\rm cubic}^{\rm PM}=0$. Using the radial-momentum correction above and retaining only the PM information through $u^7$, with $B=1-2u+\mathcal O(u^2)$ in this conversion, gives
\begin{align}
    Q_{\rm cubic}^{\rm PM}=-\Delta W_6(\gamma_0,\eta)u^6-\left[\Delta W_7(\gamma_0,\eta)-2\Delta W_6(\gamma_0,\eta)\right]u^7 .
\end{align}
By construction, the overlap-subtracted coefficients $\Delta W_6$ and $\Delta W_7$, and hence $Q_{\rm cubic}^{\rm PM}$, vanish as $\eta\rightarrow0$, ensuring that the test-mass cubic contribution already encoded in $A$ and $B$ is not double counted.

\subsection{Radiation reaction}
\label{sec:SM_EOB_dissipative}

The dissipative sector is constructed from the strong-field fluxes obtained from the modified Teukolsky calculation,
\begin{align}
    \mathcal{F}_{E}^{\mathrm{MG}}=\mathcal{F}_{E}^{\mathrm{GR,EOB}}+\delta\mathcal{F}_{E}^{\mathrm{cubic}}+\delta\mathcal{F}_{E}^{\mathrm{tide}} .
\label{eq:eob_flux}
\end{align}
The cubic and tidal corrections are taken directly from the MTF results, retaining their test-mass functional dependence, with the expansion parameters replaced by their EOB definitions. No additional finite-mass-ratio completion is introduced in the modified dissipative sector. Since the evolution begins at the reference frequency $M\Omega=20^{-3/2}$, corresponding to the upper endpoint of the MTF data in the test-mass limit, the numerically computed modified fluxes cover the inspiral interval used here without an additional weak-field post-Newtonian completion.

\subsection{Amplitude and radiative phase}

We first construct the waveform amplitude and radiative phase from the modified Teukolsky solution. Defining the complex gravitational-wave strain as $h=h_+-i h_\times$, where $h_+$ and $h_\times$ are the plus and cross polarizations, the asymptotic waveform is determined by $\Psi_{4}\sim Z_{\ell m\omega}^{\infty}e^{i\omega r_{\ast}}/r$. Here, $\Psi_4$ is defined with respect to the Hawking-Hartle tetrad adopted throughout the modified Teukolsky calculation~\cite{HawkingHartle1972, Chandrasekhar_1983, Yang:2026erj}, and therefore differs in normalization from the standard Kinnersley-tetrad quantity. The quantity $Z_{\ell m\omega}^{\infty}$ is the asymptotic Teukolsky amplitude and $r_\ast$ is the tortoise coordinate. Using the relation between $\Psi_{4}$ and the second time derivative of the strain, the amplitude and radiative phase of each mode are obtained, in the conventions adopted here, from
\begin{align}
    \lvert h\rvert=\left|\frac{Z_{\ell m\omega}^{\infty}}{2\omega^{2}}\right|,\quad \arg(h)=\arg\!\left(Z_{\ell m\omega}^{\infty}\right).
\end{align}
Together with the energy flux entering the orbital evolution, the modified Teukolsky calculation therefore provides the circular-orbit input for the dynamical phase, waveform amplitude, and radiative phase.

\subsection{Continuation through the plunge}

The modified Teukolsky calculation provides the beyond-GR radiative corrections only for circular orbits, and the numerical data used here terminate near $M\Omega=6^{-3/2}$. During the subsequent plunge the motion becomes increasingly noncircular, and a controlled treatment would require the corresponding source-driven modified-Teukolsky calculation for generic trajectories, which is not presently available. As our baseline prescription, we therefore retain the full GR EOB radiation reaction through the transition and plunge while smoothly switching off only the unavailable beyond-GR flux corrections once the circular-orbit MTF input terminates. The modified conservative dynamics is retained throughout the plunge. At the waveform level, the MTF amplitude and radiative-phase corrections are continued smoothly to the attachment time using the prescriptions described below, after which the postpeak waveform is constructed from the coupling-dependent remnant properties and modified QNM spectrum described in Sec.~\ref{sec:SM_EOB_ringdown}.

To assess the uncertainty associated with the unknown modified plunge flux, we define the total beyond-GR flux correction as $\delta\mathcal F_E\equiv\delta\mathcal{F}_{E}^{\mathrm{cubic}}+\delta\mathcal{F}_{E}^{\mathrm{tide}} $ and continue the fractional correction $\delta\mathcal F_E/\mathcal F^{\rm GR,EOB}_{E}$ from its value at the endpoint of the circular-orbit data, rather than extrapolating the absolute beyond-GR flux correction itself. We consider two alternative continuations through the plunge, in which the fractional correction is either held fixed at its endpoint value, or extrapolated according to the trend inferred from the circular-orbit data. The orbital evolution is recomputed for each case.

In the equal-mass case at $\zeta=10^{-2}$, the trend extrapolation gives the larger instantaneous fractional correction, increasing $\delta\mathcal F_E/\mathcal F^{\rm GR,EOB}_{E}$ from $\simeq 8.42\times10^{-4}$ near the endpoint of the circular sequence to about $3\%$ close to the attachment time. Nevertheless, relative to the baseline switch-off prescription introduced below, both alternatives change the complete inspiral-plunge-NQC phase by less than $8.8\times10^{-3}\,\mathrm{rad}$ and the waveform amplitude by less than about $0.17\%$. The small accumulated effect is mainly due to the short duration of the plunge, over which the enhanced flux produces only a limited additional change in the binary energy, angular momentum, and phase. We therefore use a smooth switch-off of the unknown beyond-GR flux correction as our baseline prescription, while retaining the full GR radiation reaction and all modified conservative contributions throughout the plunge. 

Introducing the rescaled variable $x=2(R/M-6)$, we define the window function
\begin{align}
    W(R)=
    \begin{cases}
    0, & R\leq 6M,\\[2pt]
    10x^3-15x^4+6x^5, & 6M<R<6.5M,\\[2pt]
    1, & R\geq 6.5M.
\end{cases}
\end{align}
The modified flux corrections used in the EOB evolution are then defined as $\delta\mathcal{F}_{\mathrm{cubic}}^{\mathrm{used}}=W(R)\delta\mathcal{F}_{\mathrm{cubic}}^{\mathrm{circ}}$ and $\delta\mathcal{F}_{\mathrm{tide}}^{\mathrm{used}}=W(R)\delta\mathcal{F}_{\mathrm{tide}}^{\mathrm{circ}}$. For the waveform amplitude, simply tapering the modified correction to zero generates an artificial peak near the transition and is therefore unsuitable. Instead, we extrapolate the ratio between the modified and GR amplitudes toward a constant value while smoothly driving its first time derivative to zero. Let $t_6$ denote the reference time defined by $M\Omega(t_6)=6^{-3/2}$, corresponding to the endpoint of the circular-orbit MTF input, and let $t_{\mathrm{attach}}$ denote the postpeak attachment time defined in Sec.~\ref{sec:SM_EOB_ringdown}. We then define the normalized plunge-time variable $\chi=(t-t_6)/(t_{\mathrm{attach}}-t_6)$. Denoting the GR and modified waveform amplitudes by $\mathcal A_{\rm GR}(t)$ and $\mathcal A_{\rm MG}(t)$, respectively, we define their ratio as $\mathcal R(t)=\mathcal A_{\rm MG}(t)/\mathcal A_{\rm GR}(t)$ and use this ratio as the quantity to be continued through the plunge. We then introduce the quintic function $S(\chi)$ to construct a smooth continuation $\mathcal R_{\rm ext}(t)$, from which the extrapolated modified amplitude $\mathcal A_{\rm MG}^{\rm ext}(t)$ is reconstructed:
\begin{align}
    S(\chi) &=1-10\chi^{3}+15\chi^{4}-6\chi^{5},\\
    \mathcal{R}_{\mathrm{ext}}(t)&=\mathcal{R}_{6}+\dot{\mathcal{R}}_{6}(t-t_{6})S(\chi),\\
    \mathcal A_{\rm MG}^{\rm ext}(t)&=\mathcal A_{\rm GR}(t)\mathcal R_{\rm ext}(t).
\end{align}
Here, $\mathcal{R}_{6}$ and $\dot{\mathcal{R}}_{6}$ denote the value and first time derivative of the amplitude ratio at $t=t_6$, respectively. This prescription matches continuously to the circular-orbit input and drives the extrapolated amplitude ratio to a constant value toward the postpeak attachment, avoiding a spurious amplitude peak during the plunge.

For the radiative phase, we continue the modified-gravity phase correction through the plunge so that the associated instantaneous-frequency correction vanishes at the postpeak attachment without introducing a net additional phase shift over the continuation interval. For $t\leq t_6$, the circular-orbit frequency correction is retained unchanged, while the continuation from $t_6$ to $t_{\rm attach}$ is constructed from its value at $t_6$. Defining
\begin{align}
    \delta\omega_{\ell m}^{\mathrm{MG}}&\equiv\frac{d}{dt}\delta\phi_{\ell m}^{\mathrm{MG}},\\
    \delta\omega_{\ell m}^{\mathrm{MG,ext}}(t)&=W_{\Phi}(t)\delta\omega_{\ell m}^{\rm MG}(t_{6})
\end{align}
for the continuation beyond $t_6$, we take
\begin{align}
    W_{\Phi}(t)&=
    \begin{cases}
    1,& t\leq t_{6},\\[2pt]
    1-40s^3+75s^4-36s^5,& t_{6}<t<t_{\mathrm{attach}},\\[2pt]
    0,& t\geq t_{\mathrm{attach}},
\end{cases}
\end{align}
with $s(t)=(t-t_{6})/(t_{\mathrm{attach}}-t_{6}).$ The extrapolated radiative-phase correction is then reconstructed by integration,
\begin{align}
    \delta\phi_{\ell m}^{\mathrm{MG,ext}}(t)&=\delta\phi_{\ell m}^{\mathrm{MG}}(t_{6})+ \int_{t_{6}}^{t}\delta\omega_{\ell m}^{\mathrm{MG,ext}}(t')dt' .
\end{align}
The window $W_{\Phi}$ is chosen so that the frequency correction vanishes smoothly at the attachment while producing no net additional phase shift over the continuation interval. In particular, $\int_0^1 W_{\Phi}(s)\,ds=0$, so that the positive and negative contributions to the phase integral cancel, yielding $\delta\phi_{\ell m}^{\mathrm{MG,ext}}(t_{\mathrm{attach}})=\delta\phi_{\ell m}^{\mathrm{MG}}(t_6)$ while simultaneously driving $\delta\omega_{\ell m}^{\mathrm{MG,ext}}$ to zero at the attachment time. This prescription provides a smooth connection between the circular-orbit waveform and the postpeak waveform without introducing an additional net radiative-phase shift during the plunge.

We also test the sensitivity to the waveform-level continuation itself while keeping the underlying EOB dynamics and flux fixed. For the amplitude, we compare the baseline continuation with a constant-ratio prescription and with an extrapolation of the trend inferred from the circular-orbit data. For the radiative phase, we compare the baseline prescription with continuations in which the instantaneous-frequency correction is either held fixed or extrapolated according to the circular-orbit trend before being smoothly driven to zero near the attachment. In the equal-mass case at $\zeta=10^{-2}$, the resulting pre-NQC differences relative to the baseline continuation satisfy $\left|\Delta\phi_{22}^{\rm cont}\right|\lesssim 2.2\times10^{-4}\ {\rm rad}$ and $\left|\Delta A_{22}/A_{22}\right|\lesssim 1.6\times10^{-3}$. These small variations again reflect the short duration of the plunge and the limited interval over which the uncertain continuation acts.

The continuation prescriptions introduced above are not unique. We choose low-order polynomial windows that satisfy the required endpoint conditions while avoiding discontinuities in the waveform and its relevant derivatives. The flux window $W$ and amplitude window $S$ are quintic smoothstep functions whose first two derivatives vanish appropriately at the endpoints. The phase window $W_{\Phi}$ is chosen to match continuously to the circular-orbit frequency correction at $t_6$, to drive both the correction and its first derivative to zero at $t_{\rm attach}$, and to satisfy $\int_0^1 W_{\Phi}(s)\,ds=0$, so that the continuation introduces no additional net radiative-phase shift. The tests described above show that reasonable variations, including deliberately aggressive extrapolations of the circular-orbit trends, lead to subdominant changes over the range $\zeta\lesssim10^{-2}$ considered here. We therefore treat the residual dependence on the plunge continuation as a modeling uncertainty of the present waveform construction, rather than interpreting the baseline prescriptions as predictions for the genuinely noncircular beyond-GR radiative corrections.

\subsection{Next-to-quasi-circular correction and calibration}\label{sec:SM_NQC}

The factorized inspiral-plunge waveform is constructed from quasi-circular radiative information and becomes progressively less accurate as the binary approaches merger. In EOB waveform models, this late-plunge behavior is corrected by a next-to-quasi-circular (NQC) factor whose coefficients are fixed by requiring the waveform near merger to reproduce NR-informed target quantities~\cite{Nagar:2018zoe,Nagar:2019wds}. For the dominant $(2,2)$ mode, we denote the inspiral-plunge waveform before the NQC correction by
\begin{align}
h_{22}^{\rm IP}(t)=A_{22}^{\rm IP}(t)e^{-i\Phi_{22}^{\rm IP}(t)},\qquad
\omega_{22}^{\rm IP}(t)\equiv\dot{\Phi}_{22}^{\rm IP}(t),
\end{align}
where $A_{22}^{\rm IP}=|h_{22}^{\rm IP}|$, $\Phi_{22}^{\rm IP}=-\arg h_{22}^{\rm IP}$, and $\omega_{22}^{\rm IP}$ are the corresponding amplitude, phase, and instantaneous frequency. The NQC-corrected waveform is written as
\begin{align}
h_{22}^{\mathrm{IP+NQC}}&=N_A e^{iN_{\Phi}} h_{22}^{\rm IP},\\
N_A&=1+a_1 n_1+a_2 n_2,\quad  N_{\Phi}=b_1 n_4+b_2 n_5.
\end{align}
Here, $N_A$ and $N_\Phi$ modify its amplitude and phase, respectively, while $a_1,a_2,b_1,b_2$ are the NQC coefficients to be determined below. The NQC basis functions are constructed from quantities that vanish or remain small in the quasi-circular limit but become increasingly important during the plunge. In the conventions adopted here, we use
\begin{align}
n_1=\left(\frac{p_{R_\ast}}{R\Omega}\right)^2,\quad
n_2=\frac{\ddot R}{R\Omega^2},\quad
n_4=\frac{p_{R_\ast}}{R\Omega},\quad
n_5=p_{R_\ast}R\Omega,
\end{align}
where $\Omega=\dot{\phi}$ is the orbital frequency. Following the standard implementation of the EOB NQC basis, $\ddot R$ is evaluated from the EOB equations of motion using $\ddot R=\dot R\,\partial_R\dot R+\dot p_{R_\ast}\,\partial_{p_{R_\ast}}\dot R$, with the explicit radiation-reaction contribution proportional to $\dot p_\phi$ omitted in this auxiliary quantity. All quantities entering this expression are evaluated along the corresponding modified EOB trajectory.

In GR EOB models, the NQC coefficients are fixed by requiring the inspiral--plunge waveform to reproduce NR-calibrated merger quantities at the NQC calibration time, which also serves as the postpeak attachment time in our construction~\cite{Nagar:2018zoe,Nagar:2019wds}. For the modified theory considered here, no corresponding NR calibration is presently available. We therefore use the GR NR-calibrated values of the waveform amplitude, amplitude derivative, instantaneous frequency, and frequency derivative as the target quantities for each modified branch, following the same strategy as the Einstein-scalar-Gauss-Bonnet (ESGB) construction of Ref.~\cite{Julie:2024fwy}. The modified dynamics and radiative inputs nevertheless enter through the branch-dependent inspiral--plunge waveform and NQC basis, and the NQC coefficients are solved anew on each modified branch by imposing these targets. The waveform sensitivity to possible deviations from these GR targets is quantified below.

For each value of the coupling, we impose the NQC conditions at $t_{\rm NQC}(\zeta) \equiv t_{\Omega_{\rm orb}}^{\rm peak}(\zeta)-M$, where $t_{\Omega_{\rm orb}}^{\rm peak}(\zeta)$ denotes the time at which the orbital frequency of the corresponding modified EOB trajectory reaches its maximum. Operationally, these target values are obtained by evaluating the GR-calibrated TEOBResumS postpeak model, described in Sec.~\ref{sec:SM_EOB_ringdown}, at the corresponding GR calibration time,  $t_{\rm NQC}^{\rm GR}=t_{\Omega_{\rm orb}}^{\rm peak,GR}-M$. Specifically, $A_{22}^{\rm target}$ and $\dot A_{22}^{\rm target}$ are the waveform amplitude and its time derivative predicted by this GR-calibrated model, while $\omega_{22}^{\rm target}$ and $\dot\omega_{22}^{\rm target}$ are its instantaneous gravitational-wave frequency and frequency derivative. In the baseline construction, these four GR values are imposed on every modified branch at its own $t_{\rm NQC}(\zeta)$,
\begin{align}
A_{22}^{\rm IP+NQC}(t_{\rm NQC}) &= A_{22}^{\rm target}, &
\dot A_{22}^{\rm IP+NQC}(t_{\rm NQC}) &= \dot A_{22}^{\rm target}, \nonumber\\
\omega_{22}^{\rm IP+NQC}(t_{\rm NQC}) &= \omega_{22}^{\rm target}, &
\dot\omega_{22}^{\rm IP+NQC}(t_{\rm NQC}) &= \dot\omega_{22}^{\rm target}.
\label{eq:SM_NQC_targets}
\end{align}
These four conditions determine the four NQC coefficients $a_1$, $a_2$, $b_1$, and $b_2$ independently for each value of $\zeta$.

We do not impose a target value for the absolute waveform phase. A common constant phase offset is arbitrary, whereas the phase difference accumulated between a modified branch and GR during the inspiral and plunge carries the accumulated relative dephasing. The NQC phase coefficients are therefore fixed by the instantaneous frequency and its derivative, while the accumulated phase is inherited continuously from the preceding inspiral--plunge evolution. Thus, the NQC correction anchors the local merger behavior to the GR-calibrated baseline without removing the accumulated modified-gravity dephasing.

To quantify the waveform sensitivity to possible beyond-GR deviations from this baseline choice, inspired by the parametrized-merger strategy of Ref.~\cite{Julie:2024fwy}, we introduce representative $O(\zeta)$ fractional variations of the amplitude and frequency target pairs,
\begin{align}
    \left(A_{22}^{\rm target},\dot A_{22}^{\rm target}\right)&\rightarrow\left(1+\zeta\kappa_A\right)\left(A_{22}^{\rm target},\dot A_{22}^{\rm target}\right),\\
    \left(\omega_{22}^{\rm target},\dot\omega_{22}^{\rm target}\right)&\rightarrow\left(1+\zeta\kappa_\omega\right)\left(\omega_{22}^{\rm target},\dot\omega_{22}^{\rm target}\right).
\end{align}
Here, $\kappa_A$ and $\kappa_\omega$ are dimensionless control parameters characterizing possible deviations from the GR-calibrated merger targets. Scaling each pair by a common factor preserves the corresponding local logarithmic slope, $\dot A_{22}^{\rm target}/A_{22}^{\rm target}$ or $\dot\omega_{22}^{\rm target}/\omega_{22}^{\rm target}$, so that the test varies the local merger amplitude or frequency scale without introducing an independent deformation of its local slope. For each variation, we recompute the NQC coefficients while keeping the underlying EOB dynamics, calibration time, remnant properties, QNM spectrum, and GR-calibrated postpeak shape parameters fixed. The resulting NQC-corrected waveform is then propagated through the postpeak construction described in Sec.~\ref{sec:SM_EOB_ringdown}, with the attachment-determined postpeak coefficients recomputed accordingly. This allows us to assess directly how such merger-target variations propagate into the final waveform.

For the equal-mass case at $\zeta=10^{-2}$, $|\kappa_A|=1$ or $|\kappa_\omega|=1$ corresponds to a $1\%$ fractional variation of the respective target pair, with the sign of $\kappa_A$ or $\kappa_\omega$ specifying the direction of the variation. A $1\%$ variation of the amplitude-target pair produces a maximum post-attachment amplitude deviation of approximately $1\%$, whereas a $1\%$ variation of the frequency-target pair leads to a maximum accumulated phase difference of about $0.0927\,\mathrm{rad}$ in the final waveform. We additionally consider $\pm3\%$ variations and find an approximately linear waveform response, with no indication of nonlinear amplification over the range tested. The stronger sensitivity to the frequency targets arises because the NQC conditions directly prescribe the local frequency evolution at the calibration point. In this control test, the QNM spectrum is held fixed, so the altered frequency boundary data are subsequently propagated into a postpeak waveform that relaxes toward the same asymptotic QNM frequency. The resulting frequency difference therefore accumulates into a phase difference, whereas the amplitude-target variation is transmitted primarily as an amplitude change without an analogous cumulative effect. Among the phenomenological inputs tested here, the GR-calibrated merger-frequency targets therefore exhibit the strongest waveform sensitivity, highlighting the importance of a future theory-specific NR calibration of the merger targets.

\subsection{Ringdown construction} \label{sec:SM_EOB_ringdown}

The postpeak waveform depends on the QNM spectrum of the remnant black hole and therefore on its final mass and spin. Since the modified dynamics changes the energy and angular momentum of the binary up to merger, these remnant quantities also receive coupling-dependent corrections. In the absence of theory-specific NR fits, we follow the prescription of Ref.~\cite{Julie:2024fwy}: the modified-minus-GR differences in the EOB energy and angular momentum at merger are added to the corresponding GR remnant values obtained from NR fits. Defining the dimensionless remnant mass and angular momentum as $m_f\equiv M_f/M$ and $j_f\equiv J_f/M^2$, where $M_f$ and $J_f$ are the mass and angular momentum of the remnant black hole, we take
\begin{align}
    m_f(\zeta)&=m_f^{\rm GR}+\frac{1}{M}\left[H_{\rm EOB}\bigl(t_{\rm mrg}(\zeta);\zeta\bigr)-H_{\rm EOB}\bigl(t_{\rm mrg}(0);0\bigr)\right],\\
    j_f(\zeta)&=j_f^{\rm GR}+\eta\left[p_\phi\bigl(t_{\rm mrg}(\zeta);\zeta\bigr)-p_\phi\bigl(t_{\rm mrg}(0);0\bigr)\right].
\end{align}
The corresponding dimensionless spin of the remnant is $\chi_f(\zeta)=J_f/M_f^2=j_f(\zeta)/m_f^2(\zeta)$. In our TEOBResumS implementation, the merger time is identified as $t_{\rm mrg}(\zeta)=t_{\Omega_{\rm orb}}^{\rm peak}(\zeta)-3M$. This prescription amounts to neglecting only the additional beyond-GR correction to the energy and angular momentum radiated after merger.

Using these remnant parameters, we construct the complex QNM frequencies from the Kerr spectrum together with the cubic-gravity corrections of Ref.~\cite{Cano:2023jbk}. The frequencies entering the waveform are
\begin{align}
    M\sigma_{22n}(\zeta)=\frac{\widehat{\sigma}_{22n}^{\rm Kerr}[\chi_f(\zeta)]}{m_f(\zeta)}+\frac{\zeta}{m_f^5(\zeta)}\delta\widehat{\sigma}_{22n}^{\rm cubic}[\chi_f(\zeta)] ,
\end{align}
where the hatted quantities are dimensionless frequencies normalized by the remnant mass $M_f$. We use the fundamental mode and first overtone, $n=0,1$, in the postpeak construction below.

Following the TEOBResumS construction~\cite{Nagar:2018zoe,Nagar:2019wds}, we model the postpeak $(2,2)$ mode as
\begin{align}
    h_{22}^{\rm post}(t)&=A_{22}^{\rm post}(t)e^{-i\Phi_{22}^{\rm post}(t)},\\
    A_{22}^{\rm post}(t)&=\left\{c_1^A\tanh\left[c_2^A(t-t_{\rm mrg})+c_3^A\right]+c_4^A\right\}e^{-\alpha_0(t-t_{\rm mrg})},\\\Phi_{22}^{\rm post}(t)&=\omega_0(t-t_{\rm mrg})+c_1^\phi\ln\left[\frac{1+c_3^\phi e^{-c_2^\phi(t-t_{\rm mrg})}+c_4^\phi e^{-2c_2^\phi(t-t_{\rm mrg})}}{1+c_3^\phi+c_4^\phi}\right]+\phi_{\rm off}.
\end{align}
Here, $t_{\rm mrg}$ denotes the branch-dependent merger time $t_{\rm mrg}(\zeta)$ defined above, $\omega_0=\operatorname{Re}\sigma_{220}$ and $\alpha_n=-\operatorname{Im}\sigma_{22n}>0$, with $c_2^A=(\alpha_1-\alpha_0)/2$ and $c_2^\phi=\alpha_1-\alpha_0$. These QNM-controlled quantities are recomputed for each value of $\zeta$ from the modified QNM spectrum obtained above. Thus, the fundamental QNM determines the late-time oscillation frequency and damping rate, while the first overtone controls the relaxation toward this asymptotic behavior. The coefficients $c_3^A$, $c_3^\phi$, and $c_4^\phi$, which determine the shape of the early postpeak waveform, are retained at their GR-calibrated TEOBResumS values. In the absence of cubic-gravity NR simulations, we do not introduce additional beyond-GR corrections to these shape parameters, following the same modeling strategy as in Ref.~\cite{Julie:2024fwy}.

The remaining coefficients $c_1^A$, $c_4^A$, $c_1^\phi$, and $\phi_{\rm off}$ are determined by attaching the postpeak waveform to the NQC-corrected inspiral--plunge waveform. The postpeak waveform is attached at the NQC calibration time, $t_{\rm attach}(\zeta)\equiv t_{\rm NQC}(\zeta)=t_{\Omega_{\rm orb}}^{\rm peak}(\zeta)-M$. Continuity of the waveform and its first time derivative at $t=t_{\rm attach}$ requires
\begin{align}
    h_{22}^{\rm post}(t_{\rm attach})&=h_{22}^{\rm IP+NQC}(t_{\rm attach}),\\
    \dot h_{22}^{\rm post}(t_{\rm attach})&=\dot h_{22}^{\rm IP+NQC}(t_{\rm attach}).
\end{align}
These two complex conditions provide four real constraints and determine $c_1^A$, $c_4^A$, $c_1^\phi$, and $\phi_{\rm off}$, ensuring a $C^1$ attachment of the two waveform pieces.

Since the relative offset between the orbital-frequency peak and the attachment time is inherited from the GR calibration, it could also receive beyond-GR corrections. We have tested an additional shift $\delta t_{\rm attach}=\zeta\kappa$ with $\kappa/M=1$; for the equal-mass case at $\zeta=10^{-2}$, the resulting changes remain below $1.55\times10^{-4}$ in relative amplitude and $2.34\times10^{-5}\,\mathrm{rad}$ in phase. This representative test indicates that an $O(\zeta M)$ correction to the attachment time produces only a subdominant waveform effect.

We also assess the sensitivity to the GR-calibrated postpeak shape parameters by varying $c_3^A$, $c_3^\phi$, and $c_4^\phi$ while keeping the inspiral--plunge waveform, NQC calibration, remnant properties, QNM spectrum, and attachment boundary data fixed. For each variation, the coefficients constrained by the $C^1$ attachment are recomputed. For the equal-mass case at $\zeta=10^{-2}$, individual $1\%$ variations of the shape parameters produce at most a $1.1\times10^{-3}$ amplitude deviation relative to the attachment amplitude and a phase difference of $5.2\times10^{-3}\,\mathrm{rad}$. Even when all three parameters are varied simultaneously by $\pm10\%$, the resulting envelope remains below $2.6\%$ in amplitude, $5.8\times10^{-2}\,\mathrm{rad}$ in phase, and $0.8\%$ in instantaneous frequency. These tests show no strong amplification of modest deviations in the GR-calibrated postpeak shape parameters.

Finally, to assess the relative importance of the two physical sources of ringdown modifications, we perform a control test analogous to that of Ref.~\cite{Julie:2024fwy}, separately including the direct cubic-gravity correction to the QNM spectrum and the correction to the remnant mass and spin. All cases are attached to the same inspiral-plunge-NQC boundary data, so that the comparison isolates the ringdown modifications. For the equal-mass case at $\zeta=10^{-2}$, the remnant correction alone changes $\operatorname{Re}\sigma_{220}$ by about $0.57\%$ and produces a maximum ringdown phase difference of $0.080\,\mathrm{rad}$, compared with about $0.50\%$ and $0.087\,\mathrm{rad}$ from the direct QNM correction alone. The direct QNM correction produces a larger change in the damping rate than the remnant correction and therefore has the larger effect on the amplitude decay. When both effects are included, the maximum ringdown phase difference reaches about $0.162\,\mathrm{rad}$. The two corrections contribute comparably to the ringdown phase, while the direct QNM correction dominates the modification of the damping rate.

Together, the coupling-dependent remnant properties and QNM spectrum, the GR-calibrated merger and postpeak inputs, and the $C^1$ attachment complete the merger--ringdown construction used in the Letter. The sensitivity tests above and in Sec.~\ref{sec:SM_NQC} show that the waveform is comparatively weakly sensitive to the tested attachment-time and postpeak-shape variations, whereas the GR-calibrated NQC merger-frequency targets exhibit the strongest waveform sensitivity among the representative phenomenological variations considered here. Separately, the remnant and direct QNM corrections provide the principal theory-specific contributions to the physical ringdown modification.

\bibliography{reference_new}